\documentclass[aps,prl,twocolumn,superscriptaddress,groupedaddress]{revtex4}
\usepackage{graphicx}  
\usepackage{dcolumn}   
\usepackage{bm}        
\usepackage{amssymb}   
\usepackage{amsmath} 
\usepackage[hang]{subfigure}
\usepackage{color}
\usepackage{comment}
\usepackage{flushend}
\usepackage{accents} 

\newcommand{\utilde}[1]{\underaccent{\scalebox{0.7}[0.7]{$\sim$}}{#1}}

\begin{document}

\title{Polarization Inversion with  $\mathcal{P}\cdot\mathcal{T}\cdot\mathcal{D}$ Symmetric Scatterers}
\author{Roee~Geva}
\affiliation{Tel Aviv University, Tel Aviv, 69978 Israel}
\author{M\'ario G. Silveirinha}
\affiliation{University of Lisbon and Instituto de Telecomunica\c{c}\~{o}es, Avenida Rovisco Pais 1, Lisboa, 1049-001 Portugal}	
\author{Raphael~Kastner}
\affiliation{Tel Aviv University, Tel Aviv, 69978 Israel}
\date{\today}

\begin{abstract}
We demonstrate, both theoretically and experimentally, that arbitrary scatterers preserving parity-time-duality ($\mathcal{P}\cdot\mathcal{T}\cdot\mathcal{D}$) symmetry inherently produce a backscattered wave whose electric field is the mirror-symmetric counterpart of the incident electric field, up to an amplitude factor, with respect to the system's characteristic mirror plane. Specifically, we establish that a general elliptically polarized wave, when reflected from such structures, exhibits a polarization state related to the polarization ellipse of the incident wave by a parity transformation. Notably, a circularly polarized wave reflects with spin angular momentum opposite to that of the incident field, in stark contrast to reflection from conventional conducting screens. These findings enable several applications such as reflective polarizers.	

\end{abstract}

\maketitle

Electromagnetic and photonic  systems that are invariant under parity ($\mathcal{P}$), time-reversal ($\mathcal{T}$), and duality ($\mathcal{D}$) can support propagation without back-scattering, even in complex environments \cite{Silveirinha_ptd, Chen2015}. This remarkable behavior stems from the fact that $\mathcal{P}\cdot\mathcal{T}\cdot\mathcal{D}$ symmetry enforces anti-symmetry in the system's scattering matrix when expressed in the $\mathcal{P}\cdot\mathcal{T}\cdot\mathcal{D}$ basis, $\mathbf{S} = -\mathbf{S}^T$. This anti-symmetry is a direct consequence of the anti-linear nature of the operator $\tilde{\mathcal{T}} = \mathcal{P}\cdot \mathcal{T} \cdot\mathcal{D}$, which behaves analogously to a fermionic time-reversal operator with $\tilde{\mathcal{T}}^2 = -\mathbf{1}$ \cite{Silveirinha_ptd, Shen2012}.

The anti-symmetric scattering matrix ensures that an electromagnetic wave incident on $\mathcal{P}\cdot\mathcal{T}\cdot\mathcal{D}$-symmetric structures cannot undergo reflection in certain propagation channels, enabling reflectionless transport in systems with an odd number of bidirectional modes \cite{Silveirinha_ptd}. This unique property opens new avenues for designing electromagnetic devices that leverage robust unidirectional propagation and efficient energy transfer \cite{Silveirinha_ptd, Chen2015, energy_sinks, Bisharat, dual_surfaces, Enrica}. Crucially, these reflectionless characteristics persist even in the presence of non-Hermitian effects, such as material dissipation \cite{Camara2024}. Additionally, $\mathcal{P}\cdot\mathcal{T}\cdot\mathcal{D}$-symmetric systems are often associated with nontrivial topological phases, further enhancing their potential for advanced electromagnetic applications \cite{Khanikaev, HePNAS, Silveirinha_ptd, sylvain_kane_mele, ct_chan_2022, Camara2024}. 

Previous studies have primarily focused on wave propagation in $\mathcal{P}\cdot\mathcal{T}\cdot\mathcal{D}$-invariant waveguides. In this Letter, we explore the scattering of waves by $\mathcal{P}\cdot\mathcal{T}\cdot\mathcal{D}$-invariant objects embedded in free space. Remarkably, we demonstrate that $\mathcal{P}\cdot\mathcal{T}\cdot\mathcal{D}$ invariance imposes unique characteristics on the scattered fields. Specifically, we show that the wave backscattered by any $\mathcal{P}\cdot\mathcal{T}\cdot\mathcal{D}$-invariant scatterer is always polarized along a direction that differs by a mirror-transformation from the incident field. We refer to this phenomenon as polarization inversion. Furthermore, we reveal that, unlike conventional materials, $\mathcal{P}\cdot\mathcal{T}\cdot\mathcal{D}$-invariant objects reverse the spin angular momentum of the wave, causing the incident and backscattered fields to rotate in opposite  directions. In addition, we demonstrate that our theory includes as a particular case the well-known Kerker condition \cite{kerker1983, fernandez2013, geffrin2012}. Rotationally symmetric non-reflective structures  \cite{Nasim}, however, are a complementary case.

$\mathcal{P}\cdot\mathcal{T}\cdot\mathcal{D}$-symmetric systems are known to eliminate reflections within a given incident mode \cite{Silveirinha_ptd}. Specifically, for a given arbitrary incident mode $\mathbf{f}^+(\mathbf{r})$ the theory of Ref. \cite{Silveirinha_ptd} guarantees that this wave cannot backscatter into the ``companion'' mode defined as
\[
\widetilde{\mathbf{f}} (\mathbf{r}) = \widetilde{\mathcal{T}}\cdot\mathbf{f}^+(\mathbf{r}'),\quad \mathbf{r}'=\mathbf{V}\cdot\mathbf{r}\]
where $\widetilde{\mathcal{T}}$ is the operator obtained by the composition of parity, time-reversal, and duality operators:
\begin{equation}\label{eq;tildeTu}
	\widetilde{\mathcal{T}}=\mathcal{P}\cdot\mathcal{T}\cdot\mathcal{D}
	=
	\begin{pmatrix}
		0&\eta_0\mathbf{V}\\
	-	\eta_0^{-1}\mathbf{V}&0
	\end{pmatrix}	\mathcal{K}.
\end{equation}
Here, $\eta_0$ is the free-space impedance, $\mathcal{K}$ is the complex conjugation operator and $\mathbf{V}$ is the $y$-coordinate inversion operator
\begin{equation}\label{eq:V_inverstion}
	\mathbf{V}
	=
	\begin{pmatrix}
		1&0&0\\
		0&-1&0\\
		0&0&1  
	\end{pmatrix}.
\end{equation}
It is implicit that the mirror plane is the $x-z$ plane. A reciprocal physical platform is invariant under the $\widetilde{\mathcal{T}}$ transformation if the material response at a certain point $\left( {x,y,z} \right)$ is related to the material response at the mirror-symmetric point $\left( {x,-y,z} \right)$ by a duality transformation. For example, in systems formed by isotropic dielectrics, the $\widetilde{\mathcal{T}}$ symmetry requires that the relative permittivity and permeability are linked as $\varepsilon \left( {x,y,z} \right) = \mu \left( {x, - y,z} \right)$.

Let us suppose first that the relevant object is a screen (e.g., an infinitely extended periodic metasurface) located in the $z=0$ plane. Additionally, we assume that the incident wave is a plane wave that illuminates the screen along the normal direction:
%
%
\begin{equation}\label{eq:pw}
	\mathbf{f}^+=
	\begin{pmatrix}	
		\mathbf{e}_0^+\\	\mathbf{h}_0^+
	\end{pmatrix}
e^{-\jmath k_0 {{\bf{\hat r}}_i}\cdot\mathbf{r}},\quad\mathbf{h}_0^+=\frac{{{\bf{\hat r}}_i}}{\eta_0}
	\times\mathbf{e}_0^+.
\end{equation}		
In the above, $k_0=\omega/c$ is the free-space wave number and ${{\bf{\hat r}}_i} = {\bf{\hat z}}$ is the direction of propagation of the incident wave. Then, from Eqs. \eqref{eq;tildeTu} and \eqref{eq:V_inverstion}, the companion mode is \cite{Silveirinha_ptd}:
%
\begin{equation}\label{eq:pwtilde}
\widetilde{\mathbf{f}}=
\begin{pmatrix}
	\tilde{\mathbf{E}}\\
	\tilde{\mathbf{H}}
\end{pmatrix}
=	\begin{pmatrix}
	\eta_0 \mathbf{V}\cdot\mathbf{h}^{+*}(\mathbf{r}')\\
	-\eta_0^{-1} \mathbf{V}\cdot\mathbf{e}^{+*}(\mathbf{r}')
\end{pmatrix} 
=\left( {\begin{array}{*{20}{c}}
{{\bf{\tilde e}}_0^{}}\\
{{\bf{\tilde h}}_0^{}}
\end{array}} \right){e^{ +\jmath{k_0}\left( {{\bf{V}} \cdot {{{\bf{\hat r}}}_i}} \right) \cdot {\bf{r}}}}
\end{equation}
Note that $\widetilde{\mathbf{f}}$ describes a plane wave that propagates along the direction $ - {\bf{V}} \cdot {{\bf{\hat r}}_i} =  - {\bf{\hat z}}$, i.e., it describes a particular mode of the reflected field with an electric field field polarized along ${\bf{\tilde e}}_0= {\bf{V}} \cdot \left( {{\bf{\hat z}} \times {\bf{e}}_0^{ + *}} \right)$. 

The $\mathcal{P}\cdot\mathcal{T}\cdot\mathcal{D}$ theory ensures that the back-scattered field ${\bf{E}}^{-}$ has a trivial projection over the companion mode, i.e., that ${\bf{\tilde e}}_0^* \cdot {{\bf{E}}^{-}} = 0$. Noting that the vectors ${\bf{\tilde e}}_0$ and ${\bf{V}} \cdot {\bf{e}}_0^ + $ are a basis of the $xoy$ plane and are orthogonal (${\left( {{\bf{V}} \cdot {\bf{e}}_0^ + } \right)^*} \cdot {\bf{\tilde e}}_0 = 0$), it is clear that the back-scattered field must be aligned with ${\bf{V}} \cdot {\bf{e}}_0^ + $, i.e., 
${{\bf{E}}^{-}} \sim {\bf{V}} \cdot {\bf{e}}_0^ + $. Thus, up to an amplitude factor, the reflected field is related to the incident field by a mirror transformation.

\begin{figure}[h!]
	\centering	
	\includegraphics[width=.8\columnwidth]{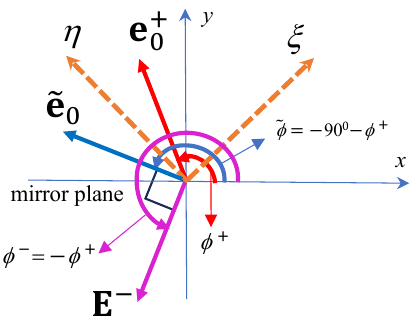}\label{fig:tautilderotationua}
	\caption{\label{fig:tautilderotationu} Effect of applying the $\mathcal{P}\cdot\mathcal{T}\cdot\mathcal{D}$ operator on a linear polarized incident polarization state $\mathbf{e}_0^+$. 
    The polarization direction is mirror-transformed by the $\mathcal{P}\cdot\mathcal{T}\cdot\mathcal{D}$ object.
    In the special cases of   $\phi^+=45^0,135^0 $ defining the coordinates $(\xi,\eta)$, the directions of $\mathbf{e}_0^+$ and $\widetilde{\mathbf{e}}_0$ coincide.
    These are the polarization ``eigenstates'' of the  $\mathcal{P}\cdot\mathcal{T}\cdot\mathcal{D}$ operation, for which the co-polarized reflection coefficients vanish. Because a $\mathcal{P}\cdot\mathcal{T}\cdot\mathcal{D}$ system does not allow back-scatter towards the $\widetilde{\phi}^+$ direction, the reflected wave $\mathbf{E}^-$ must be perpendicular to $\widetilde{\mathbf{e}}_0$. Its phase and amplitude depend on the coefficient $a$ in Eq. \eqref{eq:eimus_a}.}
\end{figure}

The geometrical relationship between the incident electric field (\({\bf{e}}_0^{+}\)), the electric field of the companion mode (\({\bf{\tilde e}}_0\)), and the backscattered field is depicted in Fig.~\ref{fig:tautilderotationu} for the case of a linearly polarized wave. Notably, the electric field of the companion mode is parallel to the mirror-transformed magnetic field of the incident wave. Furthermore, the backscattered electric field is orthogonal to the electric field of the companion mode. The incident electric field and the backscattered field, apart from a scaling factor, are related by a parity transformation with respect to the system's mirror plane (\(y=0\)).

\color{black}

In $\mathcal{P}\cdot\mathcal{T}\cdot\mathcal{D}$-symmetric systems, an incident mode ($\mathbf{f}^+$) produces no reflections in the companion mode ($\widetilde{\mathbf{f}}$). However, it may backscatter into other orthogonal modes if these are supported by the system. This is the case in our problem, where two independent propagation channels are associated with the same physical direction of propagation due to the polarization degree of freedom. In general, total suppression of backscattering can be guaranteed for a suitable excitation of the system only when the number of independent physical channels is odd \cite{Silveirinha_ptd}.

Let us write the incident electric field in terms of its components ${\bf{e}}_0^ +  = e_{0x}^ + {\bf{\hat x}} + e_{0y}^ + {\bf{\hat y}}$. Then, the backscattered electric field must be of the type
%
%
\begin{equation}\label{eq:eimus_a}
        {{\bf{E}}^ - } = a\left( {e_{0x}^ + {\bf{\hat x}} - e_{0y}^ + {\bf{\hat y}}} \right), 
\end{equation}
where $|a|\le1$ depends on the transmission and absorption levels. 
For an arbitrary incident polarization, the polarization ellipse of the reflected wave differs (apart from a scale factor) from the polarization ellipse of the incident wave by a mirror transformation with respect to the $y$-axis. In particular, for elliptically polarized waves, the absolute physical sense of rotation of the polarization ellipse is reversed compared with the sense of rotation of the incident field. In fact, one of the most remarkable features of $\mathcal{P}\cdot\mathcal{T}\cdot\mathcal{D}$-scatterers is that they flip the spin angular momentum of a wave. 

For example, suppose that the incident wave is circularly polarized to the right (RCP), corresponding to an incident field with $e_{0y}^+ = -j e_{0x}^+$. The corresponding spin angular momentum of the wave \cite{Bliokh2014, Bliokh_2015}, defined as ${\bf{\sigma}} = -j\frac{{{\bf{E}} \times {{\bf{E}}^*}}}{{{\bf{E}} \cdot {{\bf{E}}^*}}}$, is oriented along the $+z$-direction (${\bf{\sigma}}^+ = {\bf{\hat z}}$), as expected. Strikingly, upon reflection on the scatterer, the spin angular momentum becomes ${\bf{\sigma}}^- = -{\bf{\hat z}}$, which also corresponds to an RCP reflected wave. A related property (preservation of electromagnetic helicity) was previously  discussed in \cite{fernandez2013} for general dual ($\mathcal{D}$) symmetric systems.

Thus, unlike conventional mirrors (e.g., metallic or dielectric mirrors), a $\mathcal{P}\cdot\mathcal{T}\cdot\mathcal{D}$ mirror reverses the absolute sense of rotation of the reflected electric field compared to the incident field. Note that for a conventional mirror, the spin angular momentum direction is the same for both the incident and reflected waves, so that an RCP wave is reflected into a left-circularly polarized (LCP) wave, and vice versa.


Of particular interest are the cases of the ``eigenstates" $\phi^+ = 45^\circ, 135^\circ$, corresponding to a linearly polarized incident field aligned with the companion mode, i.e.,  the reflected wave $\mathbf{E}^-$ is perpendicular to $\mathbf{e}_0^+$. (see Fig.~\ref{fig:tautilderotationu}). In this case, reflection into the co-polarized mode is forbidden, meaning that the backscattered field consists exclusively of the cross-polarized wave. 
The axes of the eigenstates are labeled as $\xi$ and $\eta$ in Fig.~\ref{fig:tautilderotationu}. For an incident field polarized along $\xi$, no co-polarized reflection will be observed; however, the wave can be fully or partially reflected into the $\eta$-polarized (cross-polarized) component.

%

The described results  apply to general $\mathcal{P}\cdot\mathcal{T}\cdot\mathcal{D}$ symmetric scatterers. The scattered field for a generic object can be written as: 
\begin{equation}\label{eq:Escat}
{{\bf{E}}^{\rm{s}}}\left( {\bf{r}} \right) = {\bf{L}}\left( {{\bf{\hat r}},{{{\bf{\hat r}}}_i}} \right) \cdot {\bf{E}}_0\frac{{{e^{ - \jmath{k_0}r}}}}{{4\pi r}}. 
\end{equation}
The formula is valid in the far-field region, where the scattered wave is approximately spherical. In the above ${\bf{E}}_0$ is the incident field on the center of the object, ${{\bf{\hat r}}}$ is the observation direction, and ${{\bf{\hat r}}}_i$ is the propagation direction of the incoming plane wave. Furthermore, ${\bf{L}}\left( {{\bf{\hat r}},{{{\bf{\hat r}}}^i}} \right)$ is a matrix with units of length that determines the directional and polarization properties of the scattered field. For convenience, we define $\mathbf{E}^-({{\bf{\hat r}},{{{\bf{\hat r}}}_i}} ,\mathbf{E}_0)\equiv\mathbf{L}\left( {{\bf{\hat r}},{{{\bf{\hat r}}}_i}} \right) \cdot {\bf{E}}_0$.

The $\mathcal{P}\cdot\mathcal{T}\cdot\mathcal{D}$ symmetry requires that, for any physical channel $i$, the corresponding diagonal scattering matrix element vanishes, $S_{ii} = 0$ \cite{Silveirinha_ptd}. For incidence along ${\bf{\hat r}}_i$, the companion mode propagates along $\hat{\mathbf{r}}=- {\bf{V}} \cdot {{{\bf{\hat r}}}_i}$ and is polarized along ${\bf{V}} \cdot \left( {{{{\bf{\hat r}}}_i} \times {\bf{E}}_0^*} \right)$ (see \eqref{eq:pwtilde}). It follows that the condition $S_{ii} = 0$ imposes the requirement
$
{\left[ {{\bf{V}} \cdot \left( {{{{\bf{\hat r}}}_i} \times {\bf{E}}_0^*} \right)} \right]^*} \cdot \mathbf{E}^-= 0
$.
It is implicit here and in the formulas below that $\mathbf{E}^-$ is evaluated at $\hat{\mathbf{r}}=- {\bf{V}} \cdot {{{\bf{\hat r}}}_i}$.  Equivalently,
$
\left( {\left( {{\bf{V}} \cdot {{{\bf{\hat r}}}_i}} \right) \times \left( {{\bf{V}} \cdot {\bf{E}}_0^{}} \right)} \right) \cdot \mathbf{E}^-= 0.
$
This condition can only be satisfied if the scattered field obeys:
$
\mathbf{E}^-\sim {\bf{V}} \cdot {\bf{E}}_0.
$
In particular, when the propagation direction of the incident plane wave lies in the symmetry plane, so that ${{{\bf{\hat r}}}_i}$ is in the $y=0$ plane, it follows that the backscattered field is polarized along a direction that is mirror symmetric with respect to the incident field:
$
\mathbf{E}^-( { - {{{\bf{\hat r}}}_i},\:{{{\bf{\hat r}}}_i}},\:\mathbf{E}_0) \sim {\bf{V}} \cdot {\bf{E}}_0.
$
Thus, $\mathcal{P}\cdot\mathcal{T}\cdot\mathcal{D}$ symmetric objects inherently constrain the polarization of the backscattered field to be mirror symmetric with respect to the incident field polarization, independent of the material geometry or the detailed material response. Eigen-polarized incident field can then be defined as the case where $({\bf{V}} \cdot {\bf{E}}_0)\cdot{\bf{E}}^*_0=0$, corresponding to a vanishing co-polarized reflected wave.

In the supplementary information \cite{SM}, we present a more rigorous and general derivation of this result, which applies also to generalized non-Hermitian $\mathcal{P}\cdot\mathcal{T}\cdot\mathcal{D}$ systems that may exhibit dissipative responses.

Our theory encompasses, as a particular case, the celebrated Kerker condition \cite{kerker1983, fernandez2013, geffrin2012}, which states that spherical objects invariant under duality symmetry do not scatter in the backward direction. Indeed, for such objects, the symmetry plane can be oriented along an arbitrary direction, ensuring that both the co-polarized and cross-polarized components of the backscattered field vanish. Another interesting example of a $\mathcal{P} \cdot \mathcal{T} \cdot \mathcal{D}$-symmetric system is the class of soft-hard metasurfaces introduced by Kildal \cite{sm_kildal_0, sm_kildal_2}. 

\begin{figure}[h!]
	\includegraphics[width=.4\textwidth]{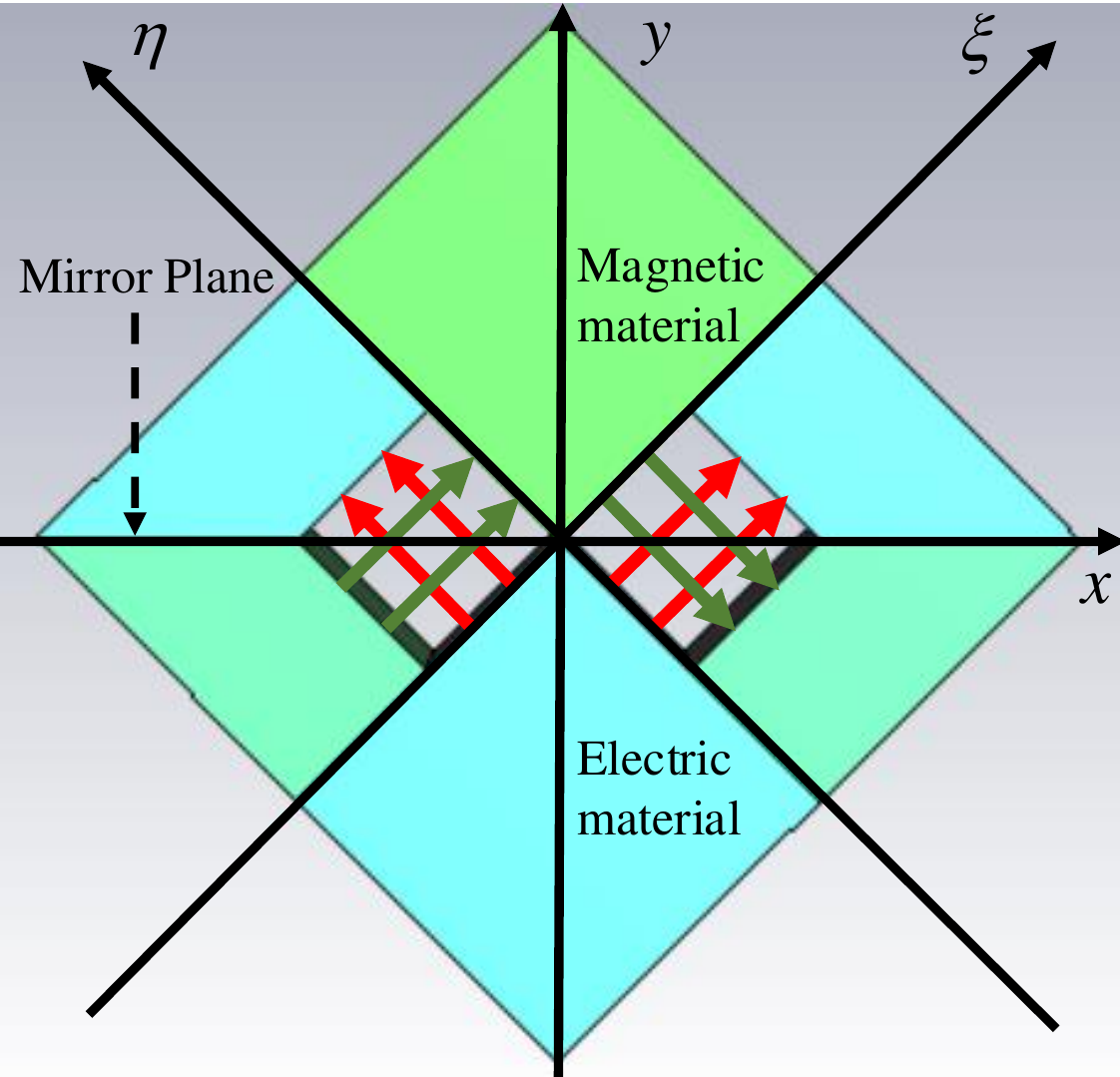}
		\caption{\label{fig:menagcli_singapore} A  $\mathcal{P}\cdot\mathcal{T}\cdot\mathcal{D}$ slab-shaped object parallel to the $x-y$ plane  is subject to plane wave illumination along the normal direction (${{{\bf{\hat r}}}_i} = {\bf{\hat z}}$)  \cite{mencagli_URSI_2021}. The electric (magnetic) materials represent idealized perfect electric (magnetic) conductors (PEC/PMC). The object has a finite thickness along the $z$-direction.       
        The electric and magnetic fields of the transmitted mode inside each hole are shown in red and green, respectively.
        The eigenpolarizations are aligned with the $\xi,\eta$ axes, determined by the directions of $45^0,135^0$  measured with respect to the $x$-axis.}
	\end{figure}

Here, we focus instead on objects with finite size, as illustrated in Fig.~\ref{fig:menagcli_singapore} \cite{mencagli_URSI_2021}. 
For convenience, we define the orientation of the incident electric field with respect to the $(\xi,\eta)$ axes. Specifically, the angle $\phi^+_\text{eigen}$ is measured relative to the $\xi$ axis.

%
%
\begin{figure}[h!]
	\subfigure[]{
		\includegraphics[width=.45\textwidth]{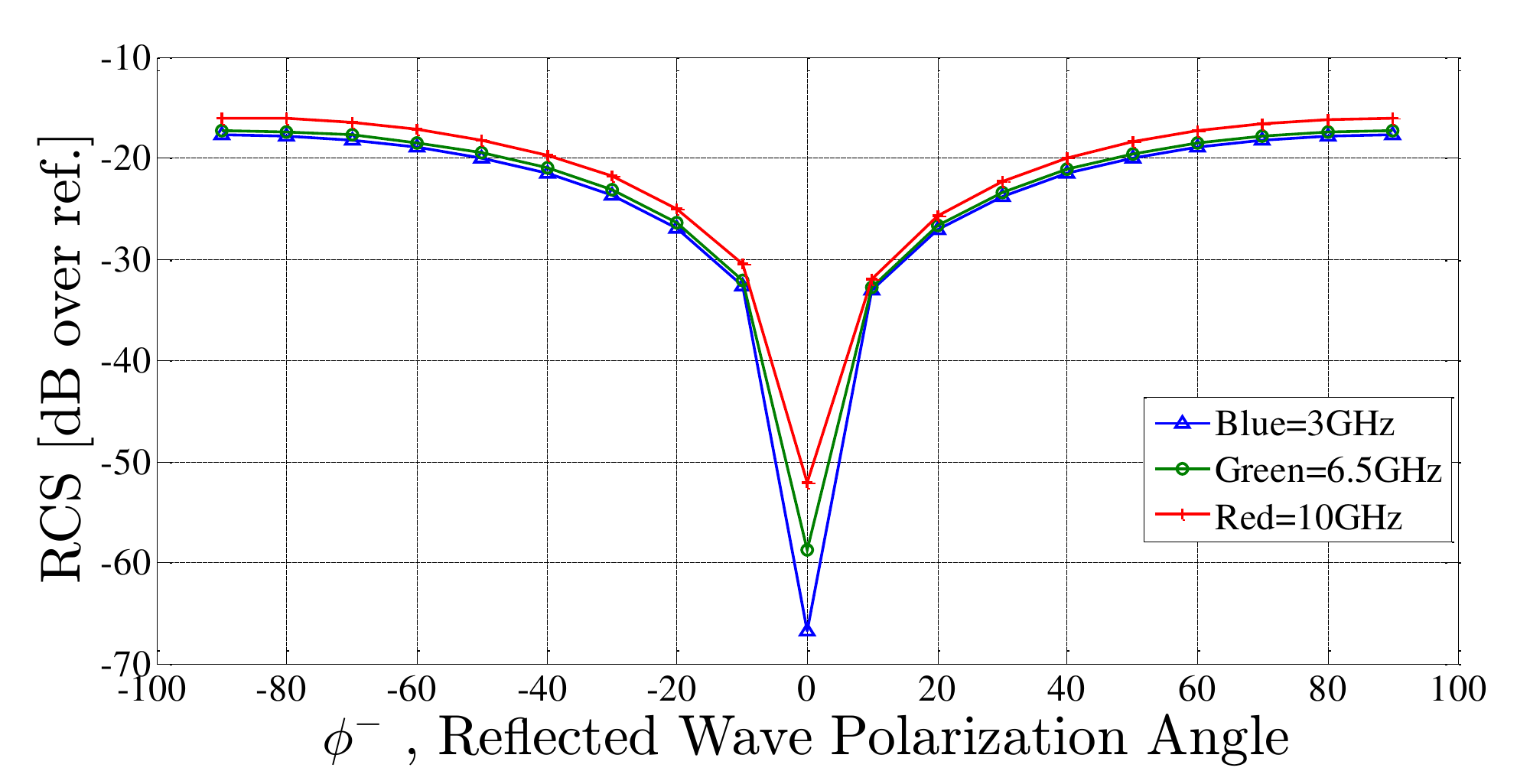}
\label{fig:no_reflection_xi}	}
\subfigure[]
{
	\includegraphics[width=.45\textwidth]{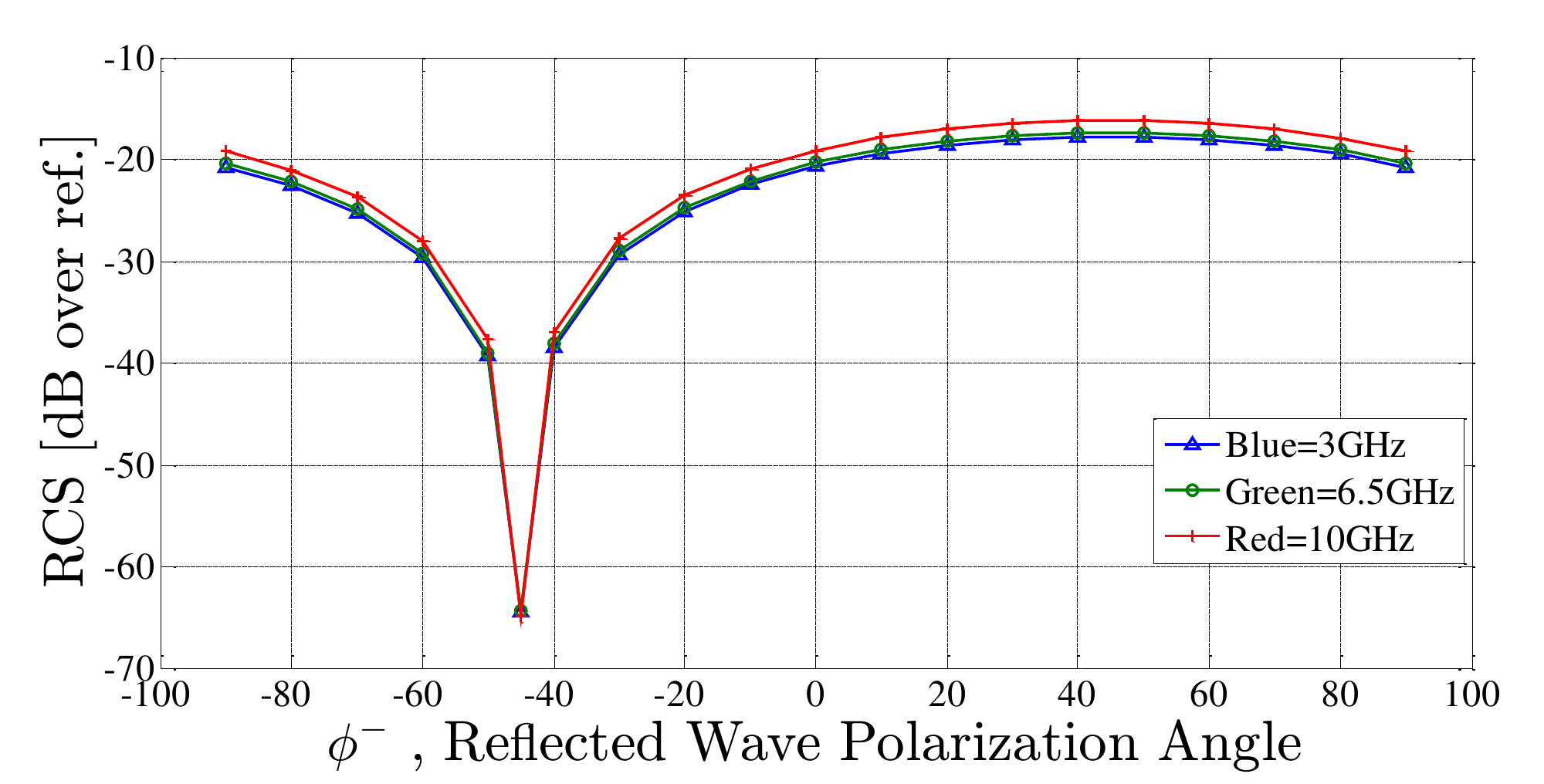}
	\label{fig:Reflection_45_deg}
}
\caption{\label{fig:no_reflection_xi_45}
Backscattered field for the object in Fig.~\ref{fig:menagcli_singapore} for normal incidence. The field amplitude, calibrated as radar cross section (RCS), is represented as a function of the orientation $\phi^-_\text{eigen}$ of the detector probe. The backscattered field is normalized to the field backscattered by a uniform PEC plate of the same size.
(a) Incident field is polarized along the eigen-polarization direction along the $\xi$ axis ($\phi^+_\text{eigen}=0$). No reflection is seen at the same polarization. Away from this point, the probe records the projection of the $\eta$ axis eigen-polarization. (b) Incident field is polarized along the $y$-axis, corresponding to $\phi^+_\text{eigen}=45^0$. The null of the back-scattered field occurs for a probe along $\phi^-_\text{eigen}=-45^0$, corresponding to the cross-polarized wave. 
    }
\end{figure}

An incident eigen-polarized wave, with its electric field oriented along $\phi^+_\text{eigen} = 0$, produces the back-scattered field shown in Fig.~\ref{fig:no_reflection_xi}. The strength of the scattered field was calibrated relative to that of a uniform perfectly electric conducting (PEC) plate of the same size. The results were obtained using the full-wave electromagnetic simulator, CST Studio Suite. No co-polarized reflection is observed, and the strongest reflection occurs for a receiving probe aligned along the cross-polarization direction, $\phi^-_\text{eigen} = \pm 90^\circ$. A similar qualitative response is observed for the other eigenpolarization, $\phi^+_\text{eigen} = 90^\circ$. The graphs in Fig.~\ref{fig:no_reflection_xi_45} simply  show the $\sin^2$-type dependence of the projection of the scattered field on any given angle.

Figure \ref{fig:Reflection_45_deg} shows the simulated back-scattered field for an incident field polarized along $\phi^+_\text{eigen} = 45^\circ$, corresponding to the $y$-axis. Since the incident field is not aligned with one of the eigen-polarizations, the reflected field exhibits a strong co-polarized component, while the cross-polarized component (along $\phi^-_\text{eigen} = -45^\circ$, corresponding to the $x$-axis and to the orientation of the companion mode field) vanishes.
 

\begin{figure}[h!]
	\includegraphics[width=.5\textwidth]{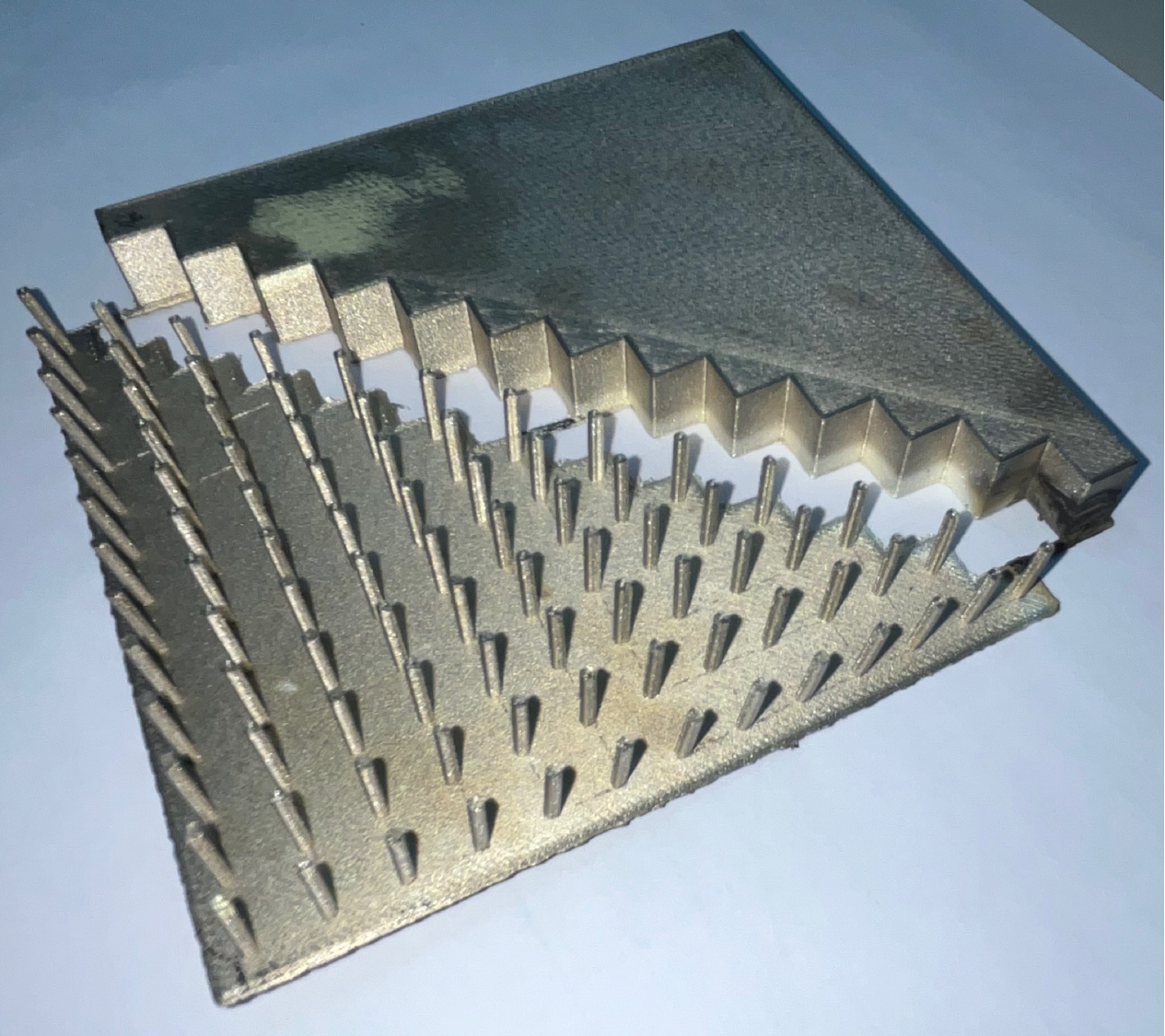}
	\caption{\label{fig:construction} Photograph of the fabricated $\mathcal{P}\cdot\mathcal{T}\cdot\mathcal{D}$ scatterer. The top-right region is metallic, providing a PEC-like response, while the bottom-left region is an artificial magnetic conductor (AMC) that behaves as a PMC at the design frequency ($f \approx 7.5~\rm{GHz}$).
    For practical reasons, the PEC and AMC regions are separated by a small air gap.
The AMC section consists of an array of square pins with a height of $11$~mm and a cross-sectional area of $1.5 \times 1.5$~mm${}^2$, arranged in a $7.5 \times 7.5$~mm${}^2$ grid. The structure was fabricated additively using a plastic substrate coated with thin layers of silver and nickel.
}
\end{figure}

To experimentally validate our theory, we designed a reflective polarizer with the geometry shown in Fig.~\ref{fig:construction}. The mirror plane ($y=0$) separates the top-right PEC-like region from the lower-left artificial magnetic conductor (PMC-like) region, which is realized using a bed-of-nails configuration \cite{King1983, Silveirinha2008, Polemi2011, Sievenpiper1999}.

The prototype was fabricated using additive manufacturing, with a 3D printer used to create a plastic structure that was subsequently coated with a thin layer of metal to achieve the desired electromagnetic properties. For comparison, a fully-PEC structure of the same dimensions was also fabricated. Both structures were tested in an anechoic chamber.

To evaluate the response of the artificial magnetic conductor (AMC), we measured the back-scattered field as a function of frequency for an incident wave polarized along the $\xi$- and $\eta$-directions. At the design frequency (7.5 GHz), where the bed of nails is expected to behave as a PMC, the co-polarization component of the reflected wave is predicted to vanish.

\begin{figure*}[h!]
 \subfigure[]{
\centering	\includegraphics[width=\columnwidth]{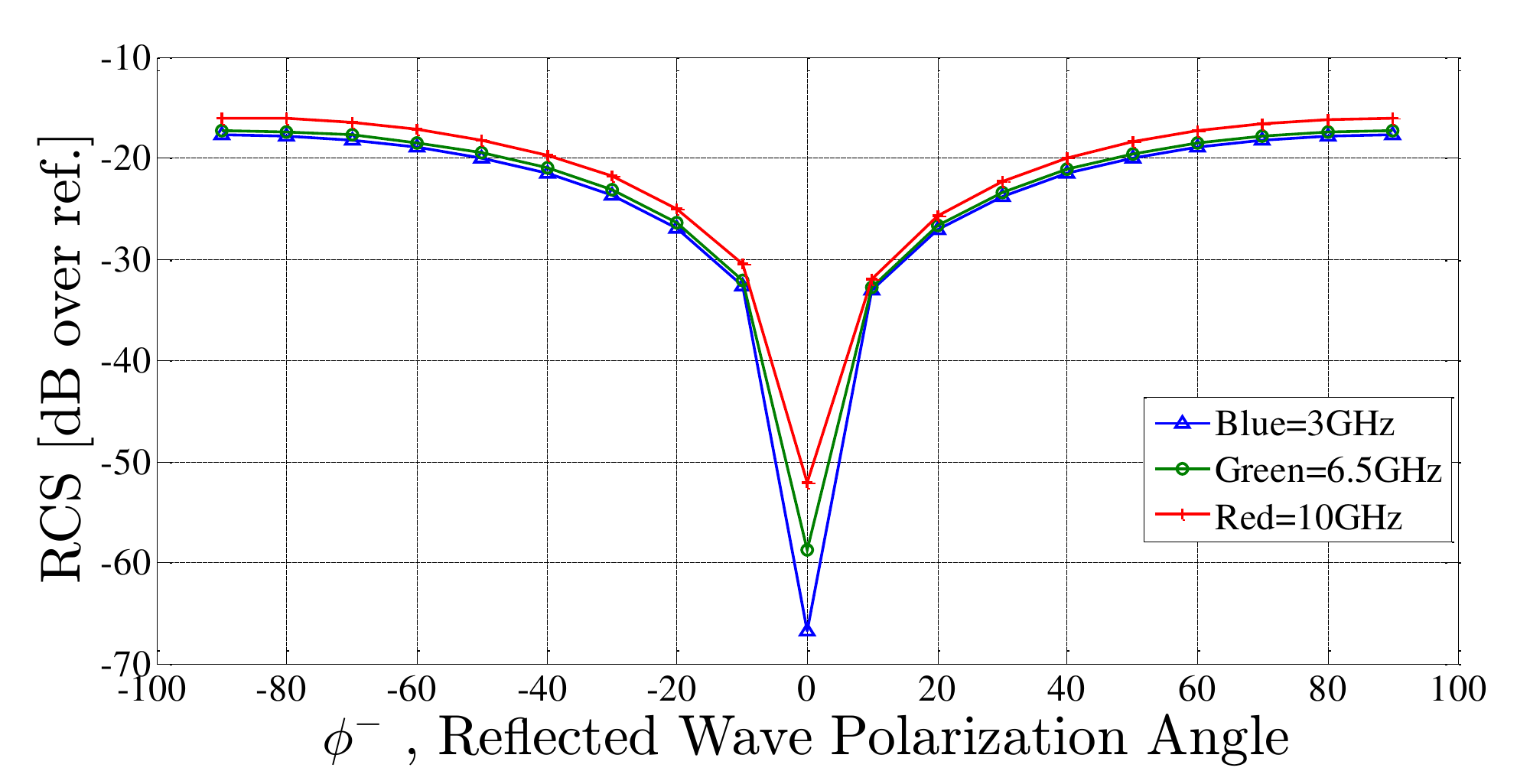}
\label{fig:graph3_az_plusminus25.pdf}  }
\subfigure[]{
	\includegraphics[width=\columnwidth]{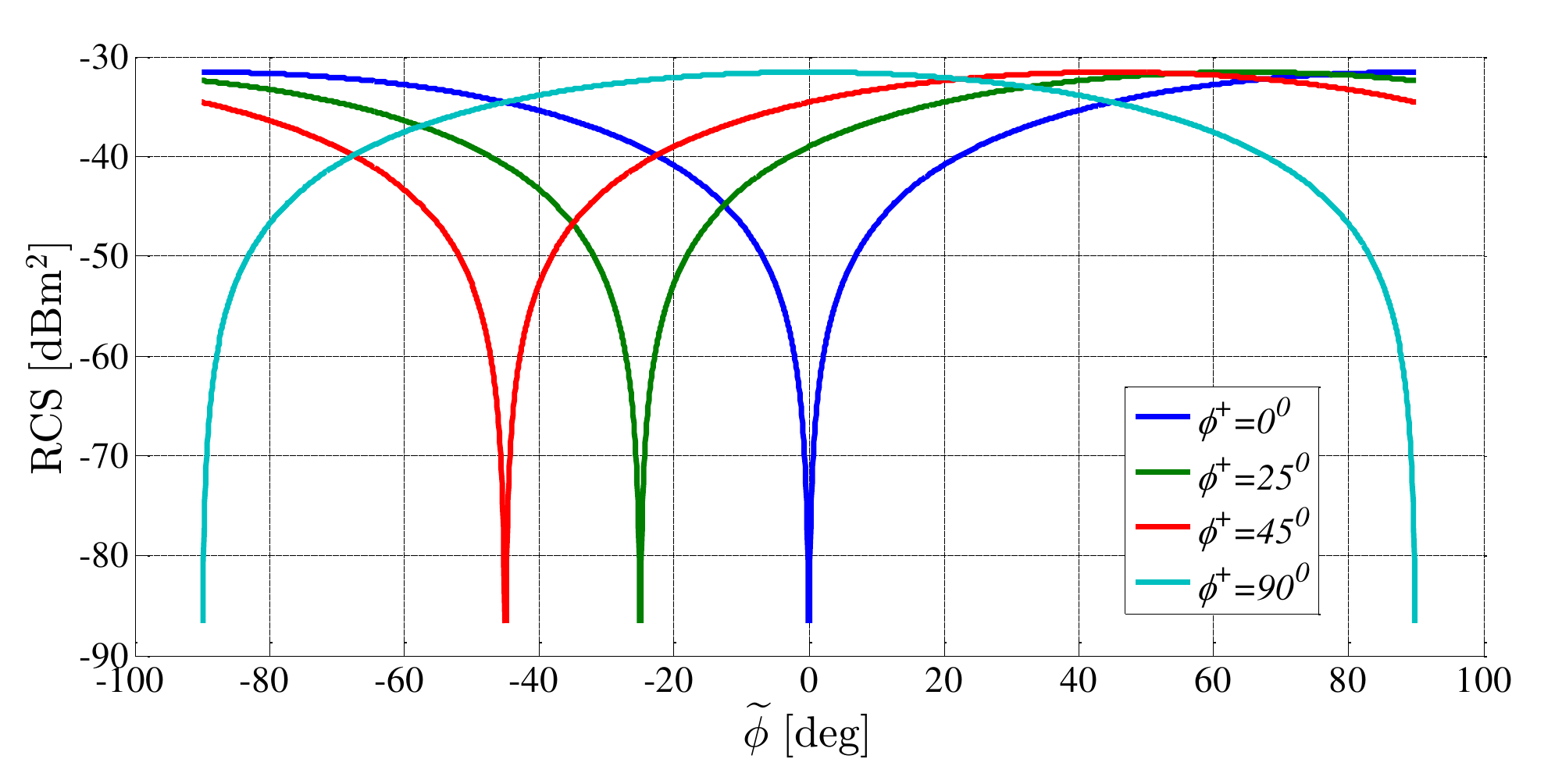}
\label{fig:graph3_az_plusminus25}
}
\subfigure[]{
\includegraphics[width=\columnwidth]{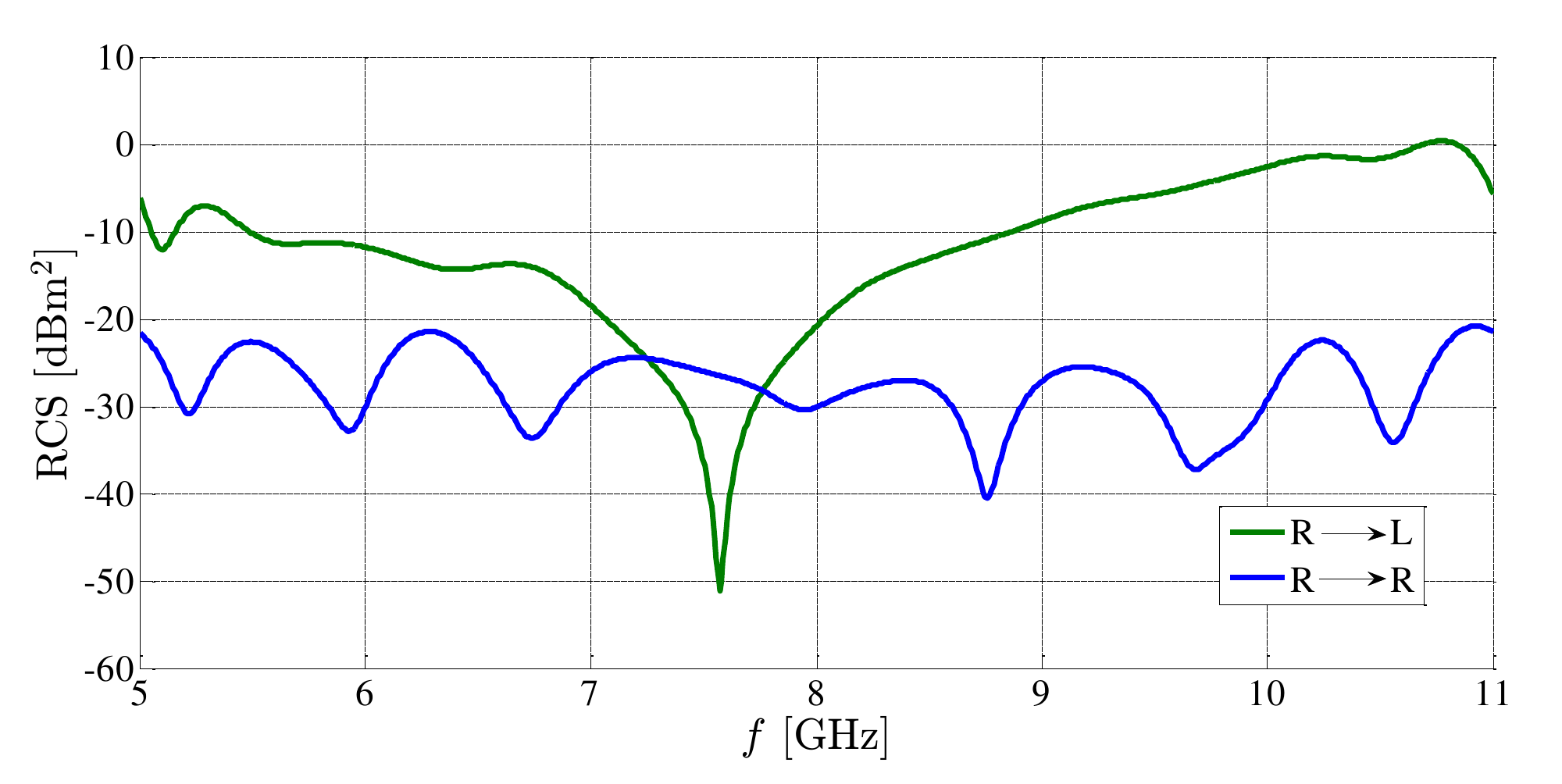}
\label{fig:experiment_circular} 
}
\caption{\label{test_results}
(a) Measured co-polarized backscattered field as a function of frequency. The incident electric field is either aligned with the $\xi$ or $\eta$ axis. At the AMC center frequency of $f=7.57$~GHz the backscattered field displays a deep null. For comparison we also depict the co-polarized backscattered field for a PEC plate of the same size with high reflection across the entire frequency range. 
(b) 
Measured backscattered field as a function of the orientation of the receiving probe for four different incident polarizations. The operational frequency is $f = 7.57$~GHz. The figure confirms that, in all cases, the backscattered electric field undergoes a mirror transformation. The measured pattern exhibits a deep null when the receiving probe is oriented parallel to the field of the companion mode.
Blue line: Incidence at the eigen-polarization along the $\xi$-axis ($\phi^+_{\rm{eigen}} = 0^\circ$). The null occurs at the same polarization as the incidence, qualitatively similar to Fig.~\ref{fig:no_reflection_xi} but for a different geometry. Green line: Incidence at $\phi^+_{\rm{eigen}} = 25^\circ$, with the null observed at $\phi^-_{\rm{eigen}} = -25^\circ$. Red line: Incidence at $\phi^+_{\rm{eigen}} = 45^\circ$, showing a null at $\phi^-_{\rm{eigen}} = -45^\circ$, qualitatively similar to Fig.~\ref{fig:Reflection_45_deg}. Cyan line: Incidence at the eigen-polarization along the $\eta$-axis ($\phi^+_{\rm{eigen}} = 90^\circ$), with the null occurring at the same polarization. All angles are defined with respect to the $\xi$-axis.
(c) Measured right- and left-circular polarization components of the backscattered field as a function of frequency for right-handed circular polarization incidence. Across most of the frequency spectrum, the scattering behavior resembles that of a PEC mirror, with dominant right-to-left polarization conversion (green curve). However, near the design frequency of the AMC, the right-to-left conversion is strongly suppressed, and the right-circular incident polarization is fully converted into a right-circular reflected polarization (blue curve), corresponding to a reversal of the spin angular momentum of the wave.
}
\end{figure*}

Figure~\ref{fig:graph3_az_plusminus25.pdf} shows the measured back-scattered field over the $5$--$11$~GHz frequency range. A pronounced null is observed at the AMC's design frequency, confirming the non-reflective property of the $\mathcal{P}\cdot\mathcal{T}\cdot\mathcal{D}$-symmetric structure for the two eigenpolarizations of the scatterer. 

The null in the co-polarized back-scattered field occurs at a frequency of $f = 7.57$~GHz. For comparison, the plot also includes the response of a reference PEC plate of the same size, which exhibits high reflection across the entire frequency range. Additionally, the numerically simulated response of the same system is provided in the supplementary information, demonstrating a qualitatively similar result \cite{SM}.


The effect of polarization inversion is experimentally confirmed in Fig.~\ref{fig:graph3_az_plusminus25}. The polarization-resolved backscattered field for both eigen-polarized and non-eigen-polarized incident fields exhibits nulls at the predicted angle $\phi^-_{\rm{eigen}} = -\phi^+_{\rm{eigen}}$, with the energy deflected into the orthogonal polarization state. These results qualitatively align with the simulations presented in Figs.~\ref{fig:no_reflection_xi}-\ref{fig:Reflection_45_deg}, which correspond to a different object.
%
%
The curve follows roughly the projection of the cross-polarized component on the measured polarization.

Figure~\ref{fig:experiment_circular} shows the experimentally measured RCP and LCP polarization-resolved components of the backscattered field as a function of frequency for an incident wave that is right-circularly polarized. As observed, across most of the frequency range, the RCP-to-LCP polarization conversion dominates (green line), consistent with the behavior of conventional metallic and dielectric mirrors, which preserve the spin angular momentum of the wave, i.e., the absolute direction of rotation of the field with respect to a fixed reference frame.

Remarkably, near the AMC design frequency ($7.5~\mathrm{GHz}$), the RCP-to-LCP polarization component exhibits a deep null. Consistent with the general theory of scattering by $\mathcal{P}\cdot\mathcal{T}\cdot\mathcal{D}$ symmetric objects, the backscattered field is dominated by the RCP-to-RCP polarization component (blue curve). This result experimentally confirms that $\mathcal{P}\cdot\mathcal{T}\cdot\mathcal{D}$ symmetric objects inherently provide a reversal of the spin angular momentum of the wave, despite being formed from fully reciprocal materials. This unique property has exciting potential applications in the design of objects with exotic scattering signatures.


In conclusion, we unveiled the unique scattering properties of $\mathcal{P}\cdot\mathcal{T}\cdot\mathcal{D}$-symmetric objects, highlighting their ability to enforce polarization inversion and reverse the spin angular momentum of backscattered waves. These results go beyond conventional scattering paradigms, revealing that the interplay of parity, time-reversal, and duality symmetries imposes strict constraints on the polarization and angular momentum of scattered fields, independent of material specifics or geometric details. 

We have experimentally validated these effects,  confirming not only the robustness of the theoretical predictions but also establishing a foundation for engineering novel photonic devices with tailored scattering signatures. By harnessing the inherent symmetry properties of $\mathcal{P}\cdot\mathcal{T}\cdot\mathcal{D}$ systems, this work opens avenues for applications in polarization control, spin-selective devices, and advanced wave manipulation technologies.
 
\begin{acknowledgments}
This work is partially supported by the Israel Science Foundation (ISF) under contract 1173/24, by the IET, by the Simons Foundation under the award 733700 (Simons Collaboration in Mathematics and Physics, ``Harnessing Universal Symmetry Concepts for Extreme Wave Phenomena''), and by FCT/MECI through national funds and when applicable co-funded EU funds under UID/50008: Instituto de Telecomunica\c{c}\~{o}es.
\end{acknowledgments}

\bibliography{article}

\appendix

\clearpage

\renewcommand{\thefigure}{S\arabic{figure}}
\setcounter{figure}{0}

\renewcommand{\theequation}{S\arabic{equation}}
\setcounter{equation}{0}

\begin{widetext}

\section{SUPPLEMENTARY INFORMATION}

In the supplementary information, we derive constraints for the directional pattern of the fields scattered by non-Hermitian $\mathcal{P} \cdot \mathcal{T} \cdot \mathcal{D}$ objects (Sects. A, B, and C). Specifically, in Sect. A, we introduce the concept of non-Hermitian $\mathcal{P} \cdot \mathcal{T} \cdot \mathcal{D}$ systems. Then, in Sect. B, we demonstrate that under a $\mathcal{P} \cdot \mathcal{D}$ transformation, a non-Hermitian $\mathcal{P} \cdot \mathcal{T} \cdot \mathcal{D}$ system is transformed into its reciprocal dual. Based on this result, we derive a generalized reciprocity relation for non-Hermitian $\mathcal{P} \cdot \mathcal{T} \cdot \mathcal{D}$ systems, which we use in Sect. C to derive a general constraint on the directional pattern of non-Hermitian $\mathcal{P} \cdot \mathcal{T} \cdot \mathcal{D}$ objects. Finally, in Sect. D, we present full-wave simulations of the back-scattered field for the experimental setup discussed in the main text.

\section{A. Non-Hermitian $\mathcal{P}\cdot\mathcal{T}\cdot\mathcal{D}$ Systems}


Let us consider a generic bianisotropic platform described by the material matrix:

\begin{equation}
{\bf{M}} = \begin{pmatrix}
{\varepsilon_0}\overline{\varepsilon} & \frac{1}{c}\bar{\xi} \\[6pt]
\frac{1}{c}\bar{\zeta} & {\mu_0}\bar{\mu}
\end{pmatrix},
\end{equation}
which links the $\bf{D}$ and $\bf{B}$ fields with the  $\bf{E}$ and $\bf{H}$ fields as $\left( {\begin{array}{*{20}{c}}
{\bf{D}}\\
{\bf{B}}
\end{array}} \right) = {\bf{M}} \cdot \left( {\begin{array}{*{20}{c}}
{\bf{E}}\\
{\bf{H}}
\end{array}} \right)$.
Following Ref.~\cite{Silveirinha_ptd}, if such a system is $\mathcal{P}\cdot\mathcal{T}\cdot\mathcal{D}$ symmetric, then the material parameters are related as:

\begin{subequations} \label{eqsm:genPTD}
\begin{align}
\overline{\varepsilon}(\mathbf{r}) &= \mathbf{V} \cdot \bar{\mu}^T(\mathbf{V}\cdot\mathbf{r}) \cdot \mathbf{V}, \\[6pt]
\bar{\zeta}(\mathbf{r}) &= -\mathbf{V} \cdot \bar{\zeta}^T(\mathbf{V}\cdot\mathbf{r}) \cdot \mathbf{V}, \\[6pt]
\bar{\xi}(\mathbf{r}) &= -\mathbf{V} \cdot \bar{\xi}^T(\mathbf{V}\cdot\mathbf{r}) \cdot \mathbf{V}.
\end{align}
\end{subequations}

In the above, the superscript ``T'' represents the transpose symmetric matrix, and the matrix $\bf{V}$ represents the parity operator. Although the above equations must hold for any $\mathcal{P}\cdot\mathcal{T}\cdot\mathcal{D}$-invariant system, they are not equivalent to $\mathcal{P}\cdot\mathcal{T}\cdot\mathcal{D}$ invariance. The latter additionally requires that ${\bf{M}} = {\bf{M}}^\dag$, i.e., that the system is conservative. 

We refer to systems that satisfy Eq.~\eqref{eqsm:genPTD} as generalized $\mathcal{P}\cdot\mathcal{T}\cdot\mathcal{D}$ platforms. These platforms are inherently non-Hermitian and can model dissipative systems, among other phenomena. A straightforward example is an isotropic material with $\varepsilon = \mu$, where both $\varepsilon$ and $\mu$ are complex-valued, representing a dissipative reciprocal $\mathcal{P}\cdot\mathcal{T}\cdot\mathcal{D}$ material. It is important to emphasize that $\mathcal{P}\cdot\mathcal{T}\cdot\mathcal{D}$ systems, in general, are not required to be reciprocal~\cite{Silveirinha_ptd}. For clarity and simplicity, all the examples in the main text focus on reciprocal platforms.
 
\section{B. Reciprocal dual of a non-Hermitian $\mathcal{P}\cdot\mathcal{T}\cdot\mathcal{D}$ system}

The reciprocal dual of a material system is another material system in which all physical parameters with odd symmetry in time (e.g., magnetic fields, velocities, etc.) are reversed. The reciprocal dual response is related to the original system response by:
\begin{equation}
{{\bf{M}}_{\rm{d}}} = {{\bf{\sigma}}_z} \cdot {{\bf{M}}^T} \cdot {{\bf{\sigma}}_z},
\end{equation}
where ${{\bf{\sigma}}_z}$ is a generalized Pauli matrix given by:
\begin{equation}
{{\bf{\sigma}}_z} = 
\begin{pmatrix}
{{\bf{1}}_{3 \times 3}} & {{\bf{0}}_{3 \times 3}} \\[6pt]
{{\bf{0}}_{3 \times 3}} & { -{{\bf{1}}_{3 \times 3}}}
\end{pmatrix}.
\end{equation}
It is important to note that the term ``dual'' in this context refers to time-reversal symmetry and is unrelated to the ``duality'' symmetry of the electromagnetic fields discussed in the main text.

The significance of the reciprocal dual concept lies in its connection to the Lorentz reciprocity theorem. Specifically, the theorem, in its most general form, states that if $\left( {\mathbf{E}, \mathbf{H}} \right)$ are generic solutions to the Maxwell equations for a system described by the matrix $\mathbf{M}$, and $\left( {\mathbf{E}_{\rm d}, \mathbf{H}_{\rm d}} \right)$ are generic solutions to the Maxwell equations for the reciprocal dual system described by the matrix $\mathbf{M}_{\rm d}$, then in a source-free region, the two field distributions satisfy:
\begin{equation} \label{eqsm:reciprocity}
\nabla \cdot \left( \mathbf{E} \times \mathbf{H}_{\rm{d}} - \mathbf{E}_{\rm{d}} \times \mathbf{H} \right) = 0.
\end{equation}

Interestingly, next we show that the reciprocal dual of a non-Hermitian $\mathcal{P}\cdot\mathcal{T}\cdot\mathcal{D}$ platform can be constructed by applying the parity-duality ($\mathcal{P}\cdot\mathcal{D}$) operator to the original system. 

For conservative $\mathcal{P}\cdot\mathcal{T}\cdot\mathcal{D}$ systems, this property can be demonstrated straightforwardly. Specifically, application of the $\mathcal{P}\cdot\mathcal{D}$ operator to the $\mathcal{P}\cdot\mathcal{T}\cdot\mathcal{D}$ operator results simply in the $\mathcal{T}$ operator, indicating that the $\mathcal{P}\cdot\mathcal{D}$ operator combined with the $\mathcal{P}\cdot\mathcal{T}\cdot\mathcal{D}$  operator transforms the original system into its time-reversed counterpart. For conservative systems, this time-reversed system coincides with the reciprocal dual. Next, we present a general proof that extends this result to dissipative platforms.

Following Ref.~\cite{Silveirinha_ptd}, a $\mathcal{P}\cdot\mathcal{D}$ transformation of the electromagnetic fields, ${\bf{f}} = {\left( {\begin{array}{*{20}{c}}
~{\bf{E}}&{\bf{H}}
\end{array}} \right)^T}$ and ${\bf{g}} = {\left( {\begin{array}{*{20}{c}}
{\bf{D}}&{\bf{B}}
\end{array}} \right)^T}$, 
is determined by:
\begin{subequations} \label{eqsm:PDop}
\begin{align}
{\bf{f}}\left( {\bf{r}} \right) \to {\utilde{\bf{f}}}\left( {\bf{r}} \right) \equiv {\cal P} \cdot {\cal D} \cdot {\bf{f}}\left( {{\bf{V}} \cdot {\bf{r}}} \right),  \\
{\bf{g}}\left( {\bf{r}} \right) \to {\utilde{\bf{g}}}\left( {\bf{r}} \right) \equiv {\cal P} \cdot \left( { - {{\cal D}^T}} \right) \cdot {\bf{g}}\left( {{\bf{V}} \cdot {\bf{r}}} \right).
\end{align}
\end{subequations} 
We use a tilde under the relevant symbol to denote a $\mathcal{P}\cdot\mathcal{D}$-transformed quantity. The $\mathcal{P}\cdot\mathcal{D}$ transformed fields can be written explicitly in terms of the original fields as:
\begin{subequations} \label{eqsm:PDfields}
\begin{align}
{\utilde{\bf{f}}}\left( {\bf{r}} \right) = \left( {\begin{array}{*{20}{c}}
0&{{\eta _0}{\bf{V}}}\\
{\eta _0^{ - 1}{\bf{V}}}&0
\end{array}} \right) \cdot {\bf{f}}\left( {{\bf{V}} \cdot {\bf{r}}} \right), \\
{\utilde{\bf{g}}}\left( {\bf{r}} \right) = \left( {\begin{array}{*{20}{c}}
0&{\eta _0^{ - 1}{\bf{V}}}\\
{{\eta _0}{\bf{V}}}&0
\end{array}} \right) \cdot {\bf{g}}\left( {{\bf{V}} \cdot {\bf{r}}} \right).
\end{align}
\end{subequations} 
Under a $\mathcal{P}\cdot\mathcal{D}$ transformation, the material matrix is transformed as:
\begin{equation}
{\bf{M}}\left( {\bf{r}} \right) \to {\utilde{\bf{M}}}\left( {\bf{r}} \right) = \left( {\begin{array}{*{20}{c}}
0&{\eta _0^{ - 1}{\bf{V}}}\\
{{\eta _0}{\bf{V}}}&0
\end{array}} \right) \cdot {\bf{M}}\left( {{\bf{V}} \cdot {\bf{r}}} \right) \cdot \left( {\begin{array}{*{20}{c}}
0&{{\eta _0}{\bf{V}}}\\
{\eta _0^{ - 1}{\bf{V}}}&0
\end{array}} \right).
\end{equation}
Straightforward calculations show that:
\begin{equation}
{\utilde{\bf{M}}}\left( {\bf{r}} \right) = \left( {\begin{array}{*{20}{c}}
{{\varepsilon _0}{\bf{V}} \cdot \bar \mu \left( {{\bf{V}} \cdot {\bf{r}}} \right) \cdot {\bf{V}}}&{\frac{1}{c}{\bf{V}} \cdot \bar \zeta \left( {{\bf{V}} \cdot {\bf{r}}} \right) \cdot {\bf{V}}}\\
{\frac{1}{c}{\bf{V}} \cdot \bar \xi \left( {{\bf{V}} \cdot {\bf{r}}} \right) \cdot {\bf{V}}}&{{\mu _0}{\bf{V}} \cdot \overline \varepsilon  \left( {{\bf{V}} \cdot {\bf{r}}} \right) \cdot {\bf{V}}}
\end{array}} \right).
\end{equation}
In particular, if the original platform is a non-Hermitian $\mathcal{P}\cdot\mathcal{T}\cdot\mathcal{D}$ system (with material parameters that satisfy Eq. \eqref{eqsm:genPTD}), then it follows that the corresponding $\mathcal{P}\cdot\mathcal{D}$ transformed system is precisely the reciprocal dual system: ${\utilde{\bf{M}}} = {{\bf{\sigma }}_z} \cdot {{\bf{M}}^T} \cdot {{\bf{\sigma }}_z} = {{\bf{M}}_{\rm{d}}}$.

Combining this result with Eq. \eqref{eqsm:reciprocity}, we conclude that if ${\bf{E'}},{\bf{H'}}$  and ${\bf{E''}},{\bf{H''}}$   are generic solutions of the Maxwell's equations in a $\mathcal{P}\cdot\mathcal{T}\cdot\mathcal{D}$ non-Hermitian platform then, in a source free-region, the following generalized reciprocity relation holds true:
\begin{equation} \label{eqsm:reciprocitygen}
\nabla  \cdot \left\{ {{\bf{E'}} \times {\utilde{\bf{H}}''} - {\utilde{\bf{E}}''} \times {\bf{H'}}} \right\} = 0\end{equation}
In the above, ${\utilde{\bf{E}}''},{\utilde{\bf{H}}''}$  are related to ${\bf{E''}},{\bf{H''}}$ through the $\mathcal{P}\cdot\mathcal{D}$ transformation [Eq. \eqref{eqsm:PDfields}].

\section{C. Scattering by non-Hermitian $\mathcal{P}\cdot\mathcal{T} \cdot\mathcal{D}$ objects}

\subsection{I. The scattering problem}

Let us consider a generic (finite sized) object standing free-space. The object is illuminated by a plane wave described by the following incident fields:
\begin{equation} \label{eqsm:inc}
{{\bf{E}}^{\rm{i}}} = {\bf{E}}_0^{\rm i}{e^{ - \jmath{k_0}{\bf{r}} \cdot {{{\bf{\hat r}}}_i}}},{\quad}{{\bf{H}}^{\rm{i}}} = \frac{1}{{{\eta _0}}}{{\bf{\hat r}}_i} \times {\bf{E}}_0^{\rm i}{e^{ -\jmath{k_0}{\bf{r}} \cdot {{{\bf{\hat r}}}_i}}}.
\end{equation}
Here, ${\bf{\hat r}}_i$ is the direction of propagation of the incident wave and ${\bf{E}}_0^{\rm{i}}$ is the complex amplitude of the wave calculated at the origin. The vector ${\bf{E}}_0^{\rm i}$ is subject to the constraint ${\bf{E}}_0^{\rm i} \cdot {\bf{\hat r}}_i = 0$.

The total fields can be decomposed into the incident and scattered components:
\begin{equation}
{\bf{E}} = {{\bf{E}}^{\rm{i}}} + {{\bf{E}}^{\rm{s}}},{\quad}{\bf{H}} = {{\bf{H}}^{\rm{i}}} + {{\bf{H}}^{\rm{s}}}.
\end{equation}

As is well known, in the far-field region  the scattered wave is spherical and is described by the following asymptotic fields ($r \to \infty$):
\begin{equation} \label{eqsm:sca}
{{\bf{E}}^{\rm{s}}} = {\bf{L}}\left( {{\bf{\hat r}},{{{\bf{\hat r}}}_i}} \right) \cdot {\bf{E}}_0^{\rm{i}}\frac{{{e^{ -\jmath{k_0}r}}}}{{4\pi r}},{\quad}{{\bf{H}}^{\rm{s}}} = \frac{1}{{{\eta _0}}}{\bf{\hat r}} \times {\bf{L}}\left( {{\bf{\hat r}},{{{\bf{\hat r}}}_i}} \right) \cdot {\bf{E}}_0^{\rm{i}}\frac{{{e^{ - \jmath{k_0}r}}}}{{4\pi r}}
\end{equation}
Here, ${\bf{\hat r}}= \left( {\cos \varphi \sin \theta ,\sin \varphi \sin \theta ,\cos \theta } \right)$ is the observation direction.
The matrix operator ${\bf{L}}\left( {{\bf{\hat r}},{{{\bf{\hat r}}}_i}} \right)$ describes the directional properties of the scattered fields, which depend on the direction of arrival of the incoming wave (${\bf{\hat r}}_i$) and on the observation direction (${\bf{\hat r}}$). As the far-fields are transverse, it is subject to the constraint ${\bf{\hat r}} \cdot {\bf{L}}\left( {{\bf{\hat r}},{{{\bf{\hat r}}}_i}} \right) = {\bf{0}}$.

It is useful to characterize how the electromagnetic fields transform under a $\mathcal{P}\cdot\mathcal{D}$ transformation. Similar to the original fields, the $\mathcal{P}\cdot\mathcal{D}$ transformed fields can also be decomposed into incident and scattered components:
\begin{equation}
{\utilde{\bf{E}}} = {\utilde{\bf{E}}^{\rm{i}}}+{\utilde{\bf{E}}^{\rm{s}}}
,{\quad}{\utilde{\bf{H}}} = {\utilde{\bf{H}}^{\rm{i}}}+{\utilde{\bf{H}}^{\rm{s}}}.
\end{equation}
With the help of Eq. \eqref{eqsm:PDfields}, it is straightforward to show that the transformed incident fields can be written explicitly as,
\begin{equation} \label{eqsm:inctilda}
{\utilde{\bf{E}}^{\rm{i}}} = {\bf{V}} \cdot \left( {{{{\bf{\hat r}}}_i} \times {\bf{E}}_0^{\rm{i}}} \right){e^{ - \jmath{k_0}{\bf{r}} \cdot \left( {{\bf{V}} \cdot {{{\bf{\hat r}}}_i}} \right)}},\quad {\utilde{\bf{H}}^{\rm{i}}} = \frac{1}{{{\eta _0}}}{\bf{V}} \cdot {\bf{E}}_0^{\rm{i}}{e^{ - \jmath{k_0}{\bf{r}} \cdot \left( {{\bf{V}} \cdot {{{\bf{\hat r}}}_i}} \right)}},
\end{equation}
whereas the transformed scattered fields are given by (in the far-field region)
\begin{equation} \label{eqsm:sctilda}
{\utilde{\bf{E}}^{\rm{s}}} =  - \left[ {{\bf{\hat r}} \times {\bf{V}} \cdot {\bf{L}}\left( {{\bf{V}} \cdot {\bf{\hat r}},{{{\bf{\hat r}}}_i}} \right) \cdot {\bf{E}}_0^{\rm{i}}} \right]\frac{{{e^{ -\jmath {k_0}r}}}}{{4\pi r}},\quad {\utilde{\bf{H}}^{\rm{s}}} = \frac{1}{{{\eta _0}}}\left[ {{\bf{V}} \cdot {\bf{L}}\left( {{\bf{V}} \cdot {\bf{\hat r}},{{{\bf{\hat r}}}_i}} \right) \cdot {\bf{E}}_0^{\rm{i}}} \right]\frac{{{e^{ - \jmath{k_0}r}}}}{{4\pi r}}.
\end{equation}
We took into account that for generic vectors $\bf{a}$ and $\bf{b}$ one has ${\bf{V}} \cdot \left( {{\bf{a}} \times {\bf{b}}} \right) =  - \left( {{\bf{V}} \cdot {\bf{a}}} \right) \times \left( {{\bf{V}} \cdot {\bf{b}}} \right)$.

\subsection{II. Constraints on the directional pattern}

In the following, we derive a constraint on the directional factor ${\bf{L}}\left( {{\bf{\hat r}},{{{\bf{\hat r}}}_i}} \right)$ for non-Hermitian $\mathcal{P}\cdot\mathcal{T} \cdot\mathcal{D}$ scatterers.

To this end, we consider two distinct plane wave illuminations of the same object, represented by the incident fields ${{\bf{E'}}^{\rm{i}}} = {\bf{E}}_0' e^{-j k_0 {\bf{r}} \cdot {\bf{\hat{r}}}'_i}$ and ${{\bf{E''}}^{\rm{i}}} = {\bf{E}}_0'' e^{-j k_0 {\bf{r}} \cdot {\bf{\hat{r}}}''_i}$, respectively. The corresponding total fields are expressed as ${\bf{E'}} = {{\bf{E'}}^{\rm{i}}} + {{\bf{E'}}^{\rm{s}}}$ and ${\bf{E''}} = {{\bf{E''}}^{\rm{i}}} + {{\bf{E''}}^{\rm{s}}}$. These fields are constrained by the generalized reciprocity relation \eqref{eqsm:reciprocitygen}, which, in its integral form, is given by:

\begin{equation}
\int\limits_{r = R} ds \; {\bf{\hat{r}}} \cdot \left( {\bf{E'}} \times \utilde{\bf{H}}'' - \utilde{\bf{E}}'' \times {\bf{H'}} \right) = 0,
\end{equation}
where the spherical integration surface is assumed to lie in the far-field region. The product of the primed and double-primed fields gives rise to three types of terms: those involving only the incident fields, those involving only the scattered fields, and cross-terms involving both incident and scattered fields. Since these terms vary according to different power laws of the distance $r$ in the far-field, they must independently satisfy the above equation. In particular, focusing on the cross-terms, we find:

\begin{equation}
\int\limits_{r = R} ds \; {\bf{\hat r}} \cdot \left( 
{{{\bf{E'}}^{\rm{i}}} \times \utilde{\bf{H}}''^{\rm{s}} - 
\utilde{\bf{E}}''^{\rm{s}} \times {\bf{H'}}^{\rm{i}} + 
{{\bf{E'}}^{\rm{s}}} \times \utilde{\bf{H}}''^{\rm{i}} - 
\utilde{\bf{E}}''^{\rm{i}} \times {\bf{H'}}^{\rm{s}}} 
\right) = 0.
\end{equation}

Let us focus on the first two terms of the integrand. Their phase variation for large $R$ (the radius of the integration surface) is controlled by the term $
{e^{-j k_0 {\bf{r}} \cdot {\bf{\hat{r}}}'_i}}{e^{-j k_0 R}} = {e^{-j k_0 R \left( 1 + {\bf{\hat r}} \cdot {\bf{\hat{r}}}'_i \right)}}.$
From the stationary phase method, the main contributions to the integral (in the $R \to \infty$ limit) arise from the directions ${\bf{\hat r}} = \pm {\bf{\hat{r}}}'_i$. It can be verified that the relevant terms vanish for the ``$+$'' sign, and hence the main contribution of the first two terms to the integral arises from the direction ${\bf{\hat r}} = -{\bf{\hat{r}}}'_i$. 
A similar argument shows that the main contribution from the last two terms of the integral arises from the observation direction ${\bf{\hat r}} = - {\bf{V}} \cdot {\bf{\hat{r}}}_i''$. Thus, the previous considerations demonstrate that the integral can vanish only if:
\begin{equation} \label{eqsm:aux1}
{{\bf{\hat r}}_i'} \cdot \left. {\left( 
{{{\bf{E'}}^{\rm{i}}} \times \utilde{\bf{H}}''^{\rm{s}} - 
\utilde{\bf{E}}''^{\rm{s}} \times {\bf{H'}}^{\rm{i}}} 
\right) } \right|_{{\bf{\hat r}} = -{\bf{\hat{r}}}'_i}
+ \left( {\bf{V}}\cdot {{\bf{\hat r}}_i''} \right)  \cdot \left. { \left( 
{ 
{{\bf{E'}}^{\rm{s}}} \times \utilde{\bf{H}}''^{\rm{i}} - 
\utilde{\bf{E}}''^{\rm{i}} \times {\bf{H'}}^{\rm{s}}} 
\right) } \right|_{{\bf{\hat r}} = - {\bf{V}} \cdot {\bf{\hat{r}}}_i''} = 0.
\end{equation}
With the help of the formulas \eqref{eqsm:inc}, \eqref{eqsm:sca}, \eqref{eqsm:inctilda} and \eqref{eqsm:sctilda}, one can show that:

\begin{subequations} 
\begin{align}
\left. {\left( 
{{{\bf{E'}}^{\rm{i}}} \times \utilde{\bf{H}}''^{\rm{s}} - 
\utilde{\bf{E}}''^{\rm{s}} \times {\bf{H'}}^{\rm{i}}} 
\right) } \right|_{{\bf{\hat r}} = -{\bf{\hat{r}}}_i'}=\frac{1}{{2\pi r{\eta _0}}}{{\bf{E}}_0'} \times \left[ {{\bf{V}} \cdot {\bf{L}}\left( { - {\bf{V}} \cdot {{{\bf{\hat r}}}_i'},{{{\bf{\hat r}}}_i''}} \right) \cdot {{{\bf{E}}}_0''}} \right], \\
\left. { \left( 
{ 
{{\bf{E'}}^{\rm{s}}} \times \utilde{\bf{H}}''^{\rm{i}} - 
\utilde{\bf{E}}''^{\rm{i}} \times {\bf{H'}}^{\rm{s}}} 
\right) } \right|_{{\bf{\hat r}} = - {\bf{V}} \cdot {\bf{\hat{r}}}_i''} =\frac{1}{{2\pi r{\eta _0}}}\left[ {{\bf{L}}\left( { - {\bf{V}} \cdot {{{\bf{\hat r}}}_i''},{{{\bf{\hat r}}}_i'}} \right) \cdot {{{\bf{E}}}_0'}} \right] \times \left( {{\bf{V}} \cdot {{{\bf{E}}}_0''}} \right).
\end{align}
\end{subequations} 
Here, we used the transverse nature of the fields and the vector identity 
$
\left( {\bf{a}} \times {\bf{b}} \right) \times \left( {\bf{a}} \times {\bf{c}} \right) = \left( {\bf{b}} \times {\bf{c}} \right),
$
which holds for generic vectors ${\bf{a}}$, ${\bf{b}}$, and ${\bf{c}}$, provided that ${\bf{a}}$ is perpendicular to both ${\bf{b}}$ and ${\bf{c}}$.

Substituting the above expressions into Eq.~\eqref{eqsm:aux1}, we arrive at the key result:
\begin{equation}
\left[ {\bf{V}} \cdot \left( {\bf{\hat r'}} \times {\bf{E}}_0' \right) \right] \cdot {\bf{L}}\left( -{\bf{V}} \cdot {\bf{\hat r'}}, {\bf{\hat r''}} \right) \cdot {\bf{E}}_0'' =  
- \left[ {\bf{V}} \cdot \left( {\bf{\hat r''}} \times {\bf{E}}_0'' \right) \right] \cdot {\bf{L}}\left( -{\bf{V}} \cdot {\bf{\hat r''}}, {\bf{\hat r'}} \right) \cdot {\bf{E}}_0'.
\end{equation}
For simplicity, we have omitted the subscript $i$ from the observation directions. This relation encapsulates the anti-symmetry property of the scattering matrix ($S_{mn} = -S_{nm}$), where the incoming wave for channel $m$ ($n$) propagates along the direction ${\bf{\hat r}}'$ (${\bf{\hat r}}''$) and the outgoing wave for channel $m$ ($n$) propagates along the direction $-\bf{V} \cdot {\bf{\hat r}}'$ (${-\bf{V} \cdot \bf{\hat r}}''$) .

In particular, if we choose the one-primed and two-primed fields to be identical, it follows that:
\begin{equation}
\left[ {\left( {\bf{\hat r}} \times {\bf{V}} \cdot {\bf{E}}_0 \right)} \right] \cdot {\bf{L}}\left( -{\bf{V}} \cdot {\bf{\hat r}}, {\bf{\hat r}} \right) \cdot {\bf{E}}_0 = 0.
\end{equation}
Since the vectors ${\bf{\hat r}} \times {\bf{V}} \cdot {\bf{E}}_0^*$ and ${\bf{V}} \cdot {\bf{E}}_0$ are orthogonal 
(${\left( {\bf{\hat r}} \times {\bf{V}} \cdot {\bf{E}}_0^* \right)^*} \cdot \left( {\bf{V}} \cdot {\bf{E}}_0 \right) = 0$), we conclude that the field scattered along the direction $-{\bf{V}} \cdot {\bf{\hat r}}$ must be aligned with a direction that is mirror-symmetric with respect to the incident field:
\begin{equation}
{\bf{L}}\left( -{\bf{V}} \cdot {\bf{\hat r}}, {\bf{\hat r}} \right) \cdot {\bf{E}}_0 \sim {\bf{V}} \cdot {\bf{E}}_0.
\end{equation}
This result is precisely the one discussed in the main text.

\section{D. Additional numerical simulations}

In this supplementary note, we present the numerically simulated back-scattered field for the fabricated prototype. The geometry of the system, as simulated in CST Studio Suite, is shown in Figure S\ref{fig:polarizer}. 

In the numerical simulation, the incident wave is assumed to be polarized along the $\xi$-direction. Figure S\ref{fig:polarizer_RCS} depicts the co-polarized back-scattered field over the $5$--$11$~GHz frequency range, as calculated using CST Studio Suite. As observed, the curve reaches a minimum at $7.42$~GHz, indicating the optimal performance of the AMC. This simulated response qualitatively aligns with the experimental results reported in the main text.

\begin{figure}[h!]
\subfigure[]{\includegraphics[width=.5\textwidth]{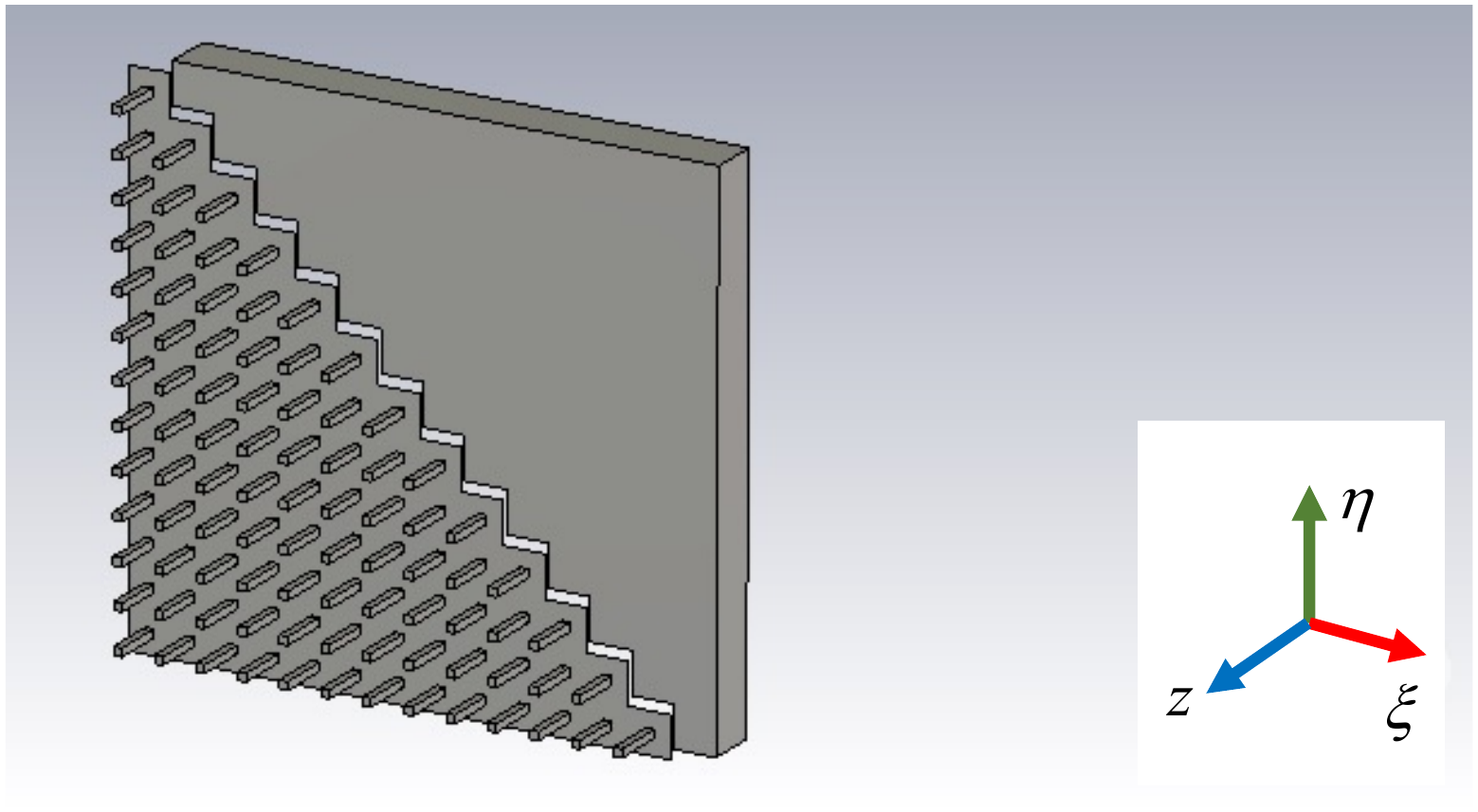}
		\label{fig:polarizer}	}
	\subfigure[]{
	\includegraphics[width=.5\textwidth]{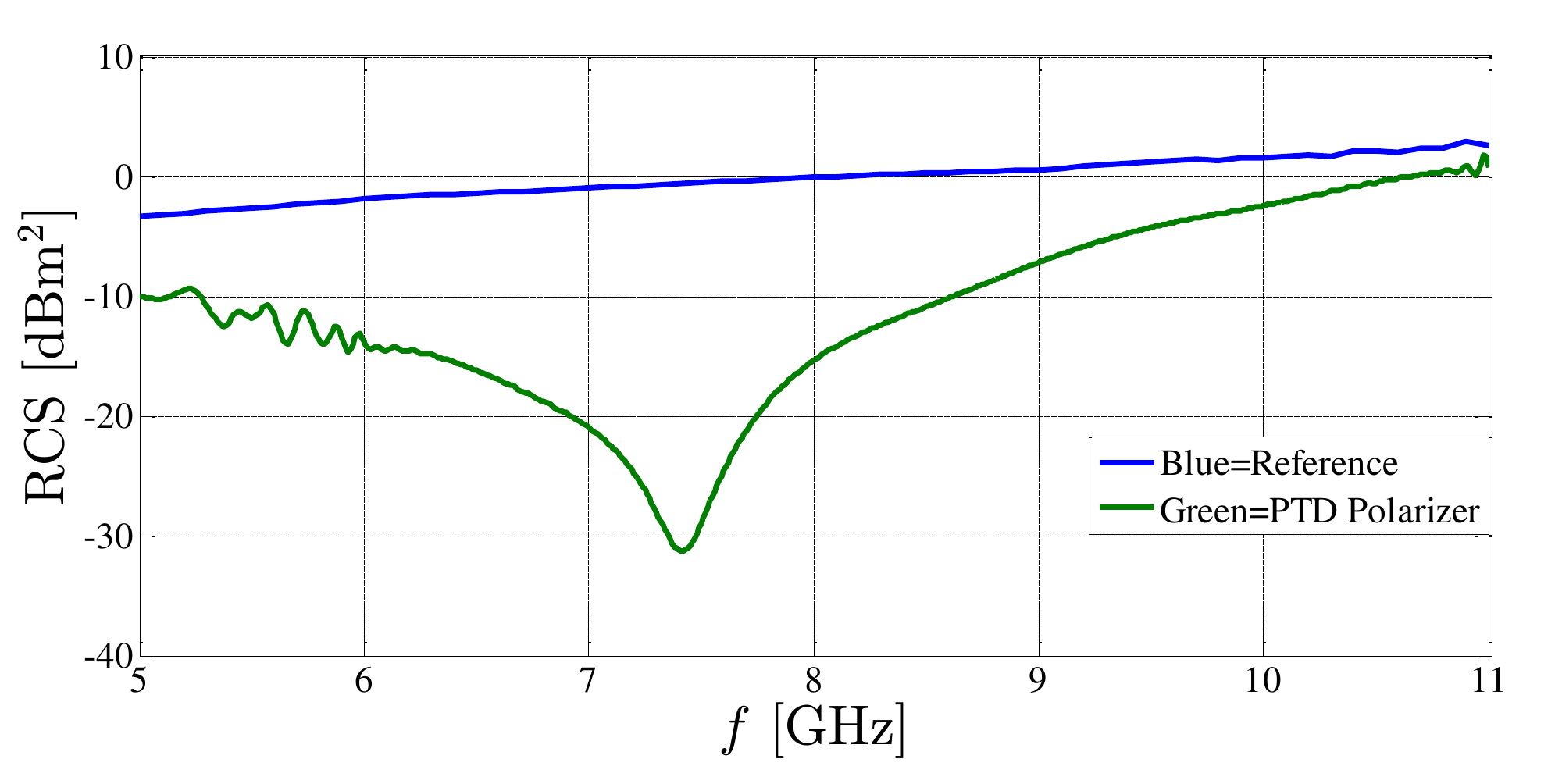}	
	\label{fig:polarizer_RCS}
}
	\caption{	\label{fig:polarizer11}
    (a) Geometry of the $\mathcal{P}\cdot\mathcal{T}\cdot\mathcal{D}$ scatterer used in the CST Studio Suite simulations. The mirror plane is slanted at $45^0$ relative to the $\xi$ or $\eta$ axes.  The top-right region is a PEC, whereas the lower-left region is an AMC, constructed as a bed-of-nails metasurface. 
    (b) Simulated co-polarized backscattered field as a function of frequency for an incident electric field oriented along the $(\xi)$ axis (green line) compared to the backscattered field for a bare conductor (blue line). The backscattered field is reduced by $-20\;$dB over a bandwidth of  $10\%$ and $10\;$dB over the bandwidth of $42\%$ around the AMC resonant frequency of $f=7.42$~GHz.
    }
\end{figure}

\end{widetext}

\end{document}